\documentclass{sig-alternate-05-2015}

\usepackage[usenames,dvipsnames,svgnames,table]{xcolor} 
\PassOptionsToPackage{hyphens}{url}
\usepackage{hyperref}
\usepackage{multirow}
\usepackage[export]{adjustbox} 
\usepackage{tabularx}
\usepackage{booktabs}
\usepackage[framemethod=tikz]{mdframed}
\newmdenv[
  innerleftmargin=7pt,
  innerrightmargin=7pt,
  tikzsetting={draw=black,dashed,line width=0.5pt,dash pattern = on 4pt off 2pt},
  linecolor=white,
  backgroundcolor=white
]{dashedbox}
\newmdenv[
  innerleftmargin=7pt,
  innerrightmargin=7pt,
  tikzsetting={draw=black, line width=0.5pt},
  linecolor=black,
  backgroundcolor=white
]{normalbox}
\newmdenv[
  topline=false,
  bottomline=false,
  rightline=false,
  skipabove=\topsep,
  skipbelow=\topsep,
  innertopmargin=0pt,
  innerbottommargin=0pt,
  innerleftmargin=7pt,
  innerrightmargin=0pt,
  tikzsetting={draw=black, line width=3pt},
  linecolor=black,
  backgroundcolor=white
]{verticalline}

\usepackage{enumitem}
\usepackage{calc}
\usepackage[export]{adjustbox} 
\usepackage{amssymb,amsmath}

\newenvironment{myquote}%
  {\list{}{\leftmargin=2em\rightmargin=2em}\item[]}%
  {\endlist}

\begin{document}


%

\CopyrightYear{2016}
\setcopyright{acmlicensed}
\conferenceinfo{ESEM '16,}{September 08 - 09, 2016, Ciudad Real, Spain}
\isbn{978-1-4503-4427-2/16/09}\acmPrice{\$15.00}
\doi{http://dx.doi.org/10.1145/2961111.2962628}

\title{Worse Than Spam: Issues In Sampling\\ Software Developers}

%
%
%
%
%

\numberofauthors{2} 
%
\author{
%
%
\alignauthor
Sebastian Baltes\\
       \affaddr{University of Trier}\\
       \affaddr{Trier, Germany}\\
       \email{research@sbaltes.com}
\alignauthor
Stephan Diehl\\
       \affaddr{University of Trier}\\
       \affaddr{Trier, Germany}\\
       \email{diehl@uni-trier.de}
}

\maketitle
\begin{abstract}
\emph{Background:} Reaching out to professional software developers is a crucial part of empirical software engineering research.
One important method to investigate the state of practice is survey research.
As drawing a random sample of professional software developers for a survey is rarely possible, researchers rely on various sampling strategies.
\emph{Objective:} In this paper, we report on our experience with different sampling strategies we employed, highlight ethical issues, and motivate the need to maintain a collection of key demographics about software developers to ease the assessment of the external validity of studies.
\emph{Method:} Our report is based on data from two studies we conducted in the past.
\emph{Results:} Contacting developers over public media proved to be the most effective and efficient sampling strategy. However, we not only describe the perspective of researchers who are interested in reaching goals like a large number of participants or a high response rate, but we also shed light onto ethical implications of different sampling strategies. 
We present one specific ethical guideline and point to debates in other research communities to start a discussion in the software engineering research community about which sampling strategies should be considered ethical.
\end{abstract}

%
%
%

\ccsdesc[500]{General and reference~Empirical studies}

%
%

%
%
\printccsdesc


\keywords{Empirical Research, Sampling, Software Developers, Ethics}

\section{Introduction}

To develop innovative ideas, processes, and tools for supporting developers in designing, writing, and maintaining software, the research community has to know their work habits and resulting needs.
Survey research in considered to be a feasible means for investigating the state of practice \cite{Cater05}.
In particular, surveys are an important empirical method used in software engineering (SE) research that can be employed to explore and describe various characteristics of a broad population \cite{Easterbrook08}.
However, reaching professional software developers with surveys is a difficult task.
Except for single companies or institutions that allow researchers to use a list of their employees, random sampling of software developers is impossible most of the time.
Researchers therefore often rely on available subjects, which is known as \textit{convenience sampling}.
Applying non-random sampling techniques like convenience sampling may lead to biased samples with limited external validity.
To mitigate the threats to external validity, researchers need detailed knowledge about the population of software developers they want to target, but this information is often not available.
Further, some of the sampling techniques that researchers employ raise ethical concerns, such as contacting developers on GitHub using email addresses users did not provide for this purpose.

In this paper, we present our research experience with different survey sampling strategies and motivate the need for a structured and systematic database with software developer demographics from different studies conducted in the past.
Such a database would enable researchers to assess the external validity of surveys conducted using non-random sampling techniques.
We further point at ethical issues that may arise with sampling approaches that researchers currently utilize and present an existing ethical guideline and works from other research communities that could inspire a discussion in the SE research community.

\section{Sampling Strategies}
\label{sec:strategies}

Generally, one can divide sampling strategies into random and non-random ones~\cite{Gravetter12, Babbie13}.
To draw a random sample, one needs an index with possible participants, which is often not available in SE research when targeting professional developers.
Therefore, many reported research findings are based on \textit{convenience samples}, which will be defined in the following.
Afterwards, we will describe our experience with different sampling strategies to recruit software developers for surveys.

\subsection{Convenience Sampling}

Often, researchers do not have access to lists of software developers, e.g., working for a particular company or in a certain area, to draw a random sample from.
Thus, it is common to rely on available subjects, which is known as \textit{convenience sampling}~\cite{Gravetter12, Babbie13} or \textit{opportunity sampling}~\cite{Searle00}.
Gravetter and Forzano~\cite{Gravetter12} describe the process of convenience sampling as ``[p]eople are selected 
on the basis of their availability and willingness to respond''.
Despite this sampling method being very popular~\cite{Gravetter12, Searle00}, it often leads to a biased sample.
One problem is that researchers are likely to approach people ``from their own social and cultural group''~\cite{Searle00}.
A specific problem of advertising online surveys, for instance using social media, is the self-selection bias~\cite{Searle00, Rosenberg08}:
Some types of people may be more likely to volunteer than others and perhaps some of them ``may be particularly keen to please'' the researcher~\cite{Searle00}.
Babbie~\cite{Babbie13} points at the limited generalizability of findings derived from a convenience sample. 
He notes that researchers ``must take care not to overgeneralize'' from such samples because convenience sampling ``does not permit any control over the representativeness''.
Further, he prompts researchers to ``alert readers to the risks associated with this method''.
Beside relying on a convenience sample and self-selection, researchers often encourage participants to advertise and share the survey, leading to a \textit{snowball sampling} approach. 
Again, this results in samples with ``questionable representativeness''~\cite{Babbie13}.

Gravetter~\cite{Gravetter12} names two strategies to mitigate problems associated with convenience sampling.
First, researchers should try to ensure that their samples are  ``reasonably representative and not strongly biased'' by carefully selecting a broad cross-section of the target population.
To do this, researchers need to know at least some basic demographic information about the population (see Section~\ref{sec:demograhics}).
Second, researchers should provide ``a clear description of how the sample was obtained and who the participants are''.
The latter is also recommended by Kitchenham et al. in the context of empirical software engineering research~\cite{Kitchenham02}.

\subsection{Experience with Sampling Strategies}

In 2013, we conducted an online survey on the use of sketches and diagrams in software engineering practice with 394 participants~\cite{Baltes14}.
For this survey, we used different sampling strategies that we are going to present in this section.
Our research was divided into four recruitment phases:
First, we recruited participants by a network of colleagues and contacts, asking them to motivate others to participate in our study.
In the second phase, we posted a call for participation in two social networks, various online communities and IRC channels.
We also contacted several German software companies and asked them to forward a call for participation to their employees.
In the third phase, we contacted a German news site for software professionals to publish a short article about our survey, asking the readers to participate.
In the last recruitment phase, we contacted people working in the area of software engineering asking them to advertise our survey on Twitter.
We also posted a call for participation in a large LinkedIn group with over 44.000 members.

\begin{figure}[t]
\centering
\includegraphics[width=\linewidth]{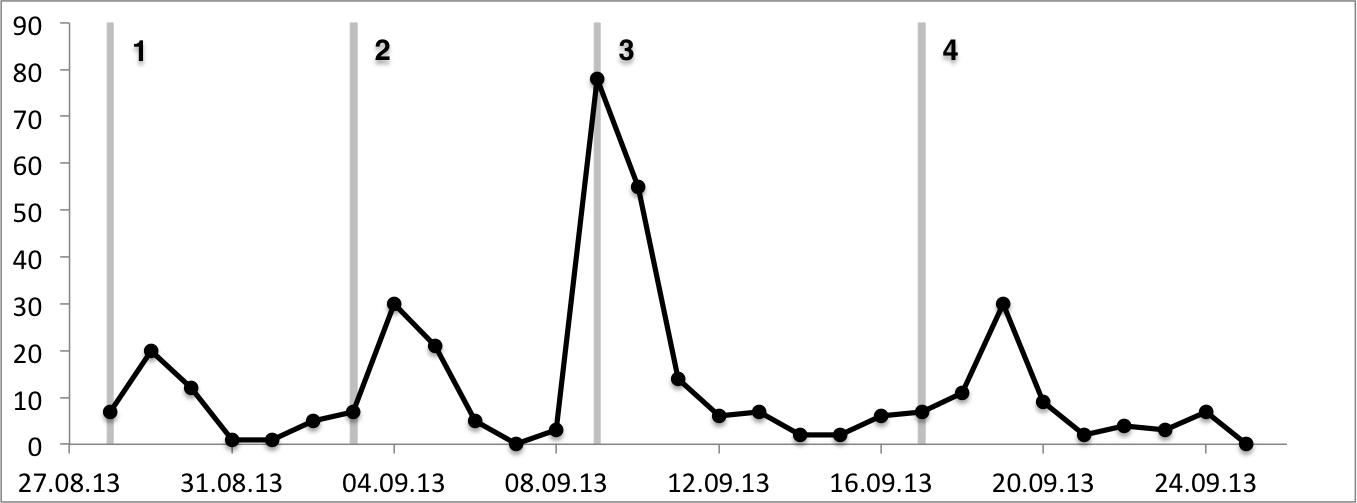}
\caption{Timeline with responses per day for the first four weeks of our online survey.}
\label{fig:timeline}
\vspace{-\baselineskip}
\end{figure}

In the following, we will report on our experience with these different ways of recruiting software developers.
Figure~\ref{fig:timeline} shows the responses we received per day during the first four weeks. The beginning of each of the four recruitment phases is highlighted.
Further, we describe our experience with sampling GitHub developers using the GHTorrent data set for a study we recently conducted.

\subsubsection{Personal Network} 

The effectiveness of using your own personal network to recruit participants for a study with software developers depends of course on the quality and quantity of the network.
As pointed out above, this approach may lead to a biased sample towards one's own background and views.
To increase the response rate, one should not only send the same email, for instance, to all members of an alumni mailing list, but formulate at least part of the email individually.
In case of our study, we were able to recruit 46 participants (12\% of total participants), before we started the next recruitment phase.
Compared to the other ones, it was the least effective sampling strategy regarding the quantity of responses.
However, it was rather efficient because contacting the own personal network does not take too much time and the contacts may forward the invitation to colleagues (snowball sampling).
In general, this strategy may be better suited for other study designs such as controlled experiments or interviews, which are not in the focus of this paper.

\subsubsection{Online Networks and Communities}

In the second phase, we posted a call for participation in two social networks, various online communities and IRC channels.
The social networks were Facebook and Google+; we used both our group's and our private accounts.
Then, we looked for online communities for software developers.
We posted calls for participation in the following communities:
\textit{DaniWeb}, 
\textit{Dev Shed}, 
\textit{dream.in.code}, 
\textit{CodeProject}, 
\textit{TopCoder}, 
\textit{Reddit}, 
and \textit{Stack Exchange's The Whiteboard}. 

Moreover, we posted on various freenode IRC channels 
including \texttt{\#\#c}, \texttt{\#\#java}, and \texttt{\#\#csharp}.
If there were answers to our posts, they were mostly positive or contained constructive feedback.
However, especially in the IRC channels, some members were very critical about using these channels to recruit participants for a survey.
Overall, this recruitment phase was the most work-intensive one.
It is hard to judge the efficiency of using online networks and communities, because we contacted software companies in parallel (see below).
Considering the fact that approximately 66 participants (17\%) answered in the second recruitment phase, which is not significantly more than in the first phase, this strategy does not seem to be more efficient.

\subsubsection{Directly Contacting Companies}

As mentioned above, in the second phase we also contacted several German software companies, which we randomly selected from online yellow pages, and asked them to forward a call for participation to their employees. 
Some of them refused to forward our request, but most of them did not answer at all.
One company, for instance, responded that they ``receive many similar requests, but [their] business is [their] priority'' and they ``cannot provide any support or working time for an interview or a questionnaire''.

This leads to a general problem when contacting people or companies to which no personal relationship exists: the lack of a \emph{gatekeeper}.
Gatekeepers are the persons who control the researcher's access to organizations \cite{Jupp06}.
The researcher needs their support to access participants inside these organizations.
In the first phase, our contacts served as gatekeepers, because they were able to ask colleagues to participate from inside their organizations.
Without such gatekeepers, it is very difficult to cross the borders of a company, especially when the researcher is conducting basic research without any immediate benefit for the company.

\subsubsection{Public Media}

In the third recruitment phase, we tried to directly reach software developers using public media.
We contacted both German and international websites, but only one German news site for software developers (heise developer, \url{http://www.heise.de/developer/}) agreed to publish a short article on our survey, asking the readers to participate.
Again, a gatekeeper in one of the editorial teams would have been very helpful.
Looking at the timeline in Figure~\ref{fig:timeline}, one can see that this article was by far the most effective recruitment channel, responsible for generating approximately 43\% of all our responses.
Considering the relatively low effort involved, in particular compared to the first two phases, we can also conclude that this was the most efficient strategy. 

\subsubsection{Using Testimonials}

In the last recruitment phase, we contacted people working in the area of software engineering asking them to advertise our survey on Twitter.
Some declined our request, but generally the success rate was higher than in the second and third phase.
In the end, three persons on Twitter with 2.300, 4.600, and 9.900 followers tweeted about our survey and we were allowed to post a call for participation on one large LinkedIn group with over 44.000 members, focusing on software architecture.
It is hard to judge how many answers were actually generated by the last phase, as it overlaps with the prior ones, but we can estimate the last phase to have generated about 73 answers (19\%).

\subsubsection{Using GHTorrent}
\label{sec:using-ghtorrent}

For a later yet unpublished study, we employed the GHTorrent data set for sampling software developers.
GHTorrent is a project collecting data about all public projects available on GitHub, providing this data via MongoDB and MySQL both online and as data dumps~\cite{Gousios13}.
In a recent meta-analysis, Cosentino et al. found that GHTorrent was the most commonly used data source for research on GitHub~\cite{Cosentino16}.
One table in the relational database scheme of GHTorrent provides information about all users who were active in the monitored time frame. 
Information provided includes the GitHub username, real name, company, location, and email address; the latter was only present until March 2016 (see Section~\ref{sec:ethical-considerations}).
Not every field is available for every user, but the data quality is quite good.
By combining the user table with other available information in the data set like commits, issue comments, or pull request, it is possible to get a good overview of the activities of a user.
In our case, we used this information to identify active Java developers and contacted a random sample of them via email, asking them to participate in our survey.

The possibility to draw a random sample from a set of developers possessing certain characteristics is very compelling and normally not possible outside a single company or institution.
Thus, there have been several research papers over the last years following this approach.
However, certain ethical issues arise, which will be discussed in Section~\ref{sec:ethical-considerations}.

\subsubsection{Other Sampling Strategies}

Beside the sampling strategies described above, there exist further strategies like using commercial recruiting services like \textit{Survata} 
or crowdsourcing platforms like \textit{Amazon Mechanical Turk}. 
However, it is questionable if these services are suited for reaching professional software developers.
Besides, researchers may personally advertise their surveys at industry conferences, which is rather time-consuming, or use students as participants, which is again not well-suited if professional software developers are the target population.
When conducting research with students as participants, special methodological and ethical issues arise~\cite{Wright06}, which are out of scope for this paper as we concentrate on professional software developers. 

\subsubsection{Summary}

We reported on our experience with different ways of recruiting software developers for an online survey.
Of course the response rates may also depend on the studied problem and the potential benefit for the participants.
However, as we collected about 43\% of the responses in one of the four phases, our recommendation is still valid.  
In the course of this paper, we will highlight our results and propositions for the presented issues as follows:

\begin{verticalline}
\textbf{Issue:} Reaching professional software developers with surveys can be a cumbersome and time-consuming task.\\
\textbf{Experience:} 
For us, convincing a news site for software developers to write about our survey was the most efficient and effective recruiting channel.
In addition to that, finding ``testimonials'' or gatekeepers in companies or social networks is very important to cross (company) boundaries, build trust, and reach many developers.
\end{verticalline}

\begin{verticalline}
\textbf{Issue:}  Commonly employed sampling approaches lead to convenience samples with several issues and biases.\\
\textbf{Experience:}  One sampling approach we employed did not result in a convenience sample, namely using the GH\-Torrent data set to randomly draw a sample from all monitored Java developers on GitHub.
However, ethical issues exist with this sampling approach (see Section~\ref{sec:ethics-sampling}).
Generally, not all possible sampling issues and biases can be mitigated, but it is important that researchers are aware of limitations and openly communicate them.
\end{verticalline}

One strategy to deal with convenience samples is to collect and present demographic data about participants to allow others to compare the sample at hand with what is know about the population.
In the next section, we present possible data sets to compare samples to, but we also point at the problem that no structured and systematic source for software developer demographics exists.  

\section{Demographics}
\label{sec:demograhics}

\begin{figure}[t]
\centering
\includegraphics[width=\linewidth, trim=0.3in 0.4in 0.0in 0.0in, clip=true]{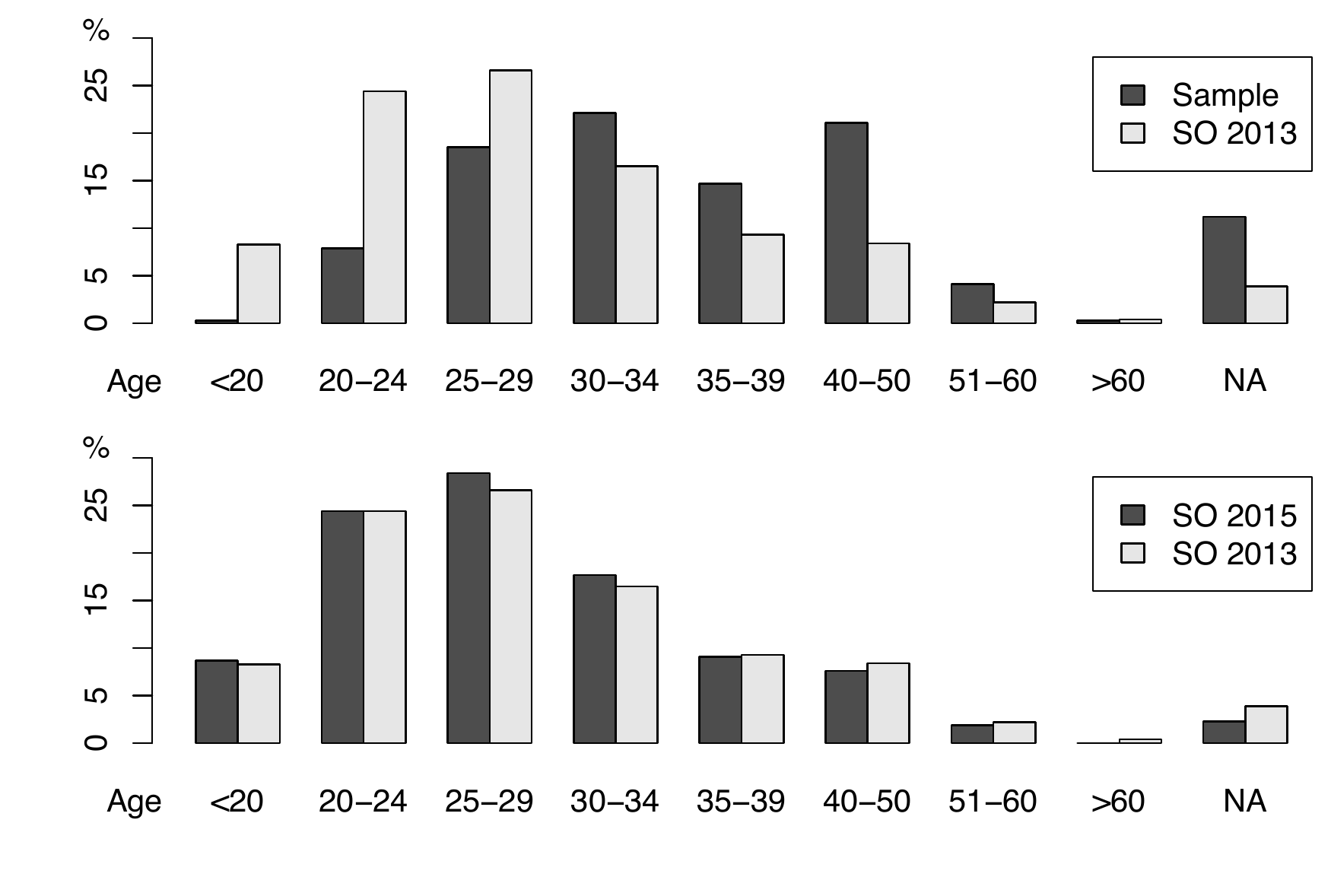}
\caption{Age distribution of our sample and the Stack Overflow Developer Survey 2013 and 2015.}
\label{fig:so-survey}
\vspace{-\baselineskip}
\end{figure}

As motivated above, one way of dealing with convenience samples is to describe the sampled population as thoroughly as possible to be able to compare it to other sample populations.
Unfortunately, there is currently no structured and systematic source where demographics from different studies involving software developers are collected.
Probably the best resource available at the moment are the results of the yearly Stack Overflow developer surveys.
The complete data sets with all responses for the surveys from 2010, 2011, 2012, 2013, and 2015 are available online.
In the 2015 survey, for instance, 26,086 developers participated  (see \cite{StackExchange15}).

To compare the sample from our study about sketches and diagrams in practice~\cite{Baltes14} to the Stack Overflow (SO) data, we chose the 2013 data set with 7,643 responses, because our data was collected in the same year.
Both our sample and the SO data set contain information about participants' age, gender, and experience in software development.
After adjusting the scales we were able to compare these three demographics and found that our sample was, compared to the SO data set, biased towards older and more experienced developers (see Figure~\ref{fig:so-survey} for the age comparison).
Further, in our study more participants refused to provide their age (5.6\% vs.~1.8\%) and we had fewer female respondents (2.8\% vs.~4.8\%).
We also compared the 2013 SO data set to the 2015 SO data set (n=26,086) and found no major differences in participants' basic demographics (see Figure~\ref{fig:so-survey} for the age comparison).
The SO data sets also provide information about participants' occupation, which allows researchers to filter the data.
One could, for instance, compare a particular sample only to developers in the SO data set who identified themselves as ``desktop developers'' or other roles depending on the context of one's own study.

With the above comparison example, we want to motivate how researchers could profit from a large database with demographic information from different studies.
It is important to have access to a diverse selection of data sets, as data from single websites or services like SO may likely be biased in certain ways, regardless of a large sample size~\cite{Hargittai15}.
There may be a significant difference in software developers who are active on Stack Overflow or GitHub compared to software developers not using such websites or services.

Beside knowledge of demographics, researchers surveying users of single services like GitHub would benefit from information about typical response rates for such surveys, as a low response rate may lead to nonresponse bias~\cite{Sax03}.
Currently, there are only few rather old papers that describe typical response rates for software engineering surveys~\cite{Cater05, Punter03}.
When describing survey samples, another important aspect is sample size.
Having a database with information about SE surveys conducted in the past would enable researchers to determine the local standard for samples sizes in the SE research community, similar to what Caine has done for the CHI community~\cite{Caine16}.

\begin{verticalline}
\textbf{Issue:} Relying on convenience samples or samples drawn from single online platforms may lead to biased results.\\
\textbf{Proposition:}
Thoroughly describing a study sample and comparing it to other samples is an important strategy to address this issue.
However, a structured and systematic source with key demographics, samples sizes, and response rates for surveys conducted by the software engineering research community does not exist.
We should build and maintain such a database to be able to compare samples and derive local standards.
\end{verticalline}

\section{Ethical Considerations}
\label{sec:ethical-considerations}

Ethics are ``rules of behavior based on ideas about what is morally good and bad''~\cite{Merriam16}.
The sampling strategies described above not only differ in terms of their effectiveness and efficiency regarding the number of software developers that can be reached, but they also differ regarding their ethical implications.
Researchers should be aware that contacting software developers causes costs even if the contacted individual does not decide to participate in the survey.
Reading the invitation email and deleting it also takes time.
We got alerted by a response of one software developer whom we contacted via the email address on his GitHub profile:
 
\begin{myquote}
\itshape
``I get emails like this every week. You might not realize this but it's majorly annoying and I consider this problem now worse than spam, since Google at least filters out spam for me. [...] [Y]ou send one, I get one per week -- or more.
I was playing along for the first 30 or so, and by now (after several hundred emails) I'm quite annoyed.''
\end{myquote}

Survey invitations being perceived as spam is not only an ethical challenge~\cite{Shilton16}, but also a problem for the resulting sample.
If certain very active users get contacted very often by researchers, it becomes less likely that they respond to such survey requests, resulting in a selection bias towards people who were contacted less often in the past.
In the following, we will present general ethical principles and a concrete code of conduct that deals with issues in sampling participants via email.
We concentrate on ethics and consider the legal situation to be out of the scope of this paper.

\subsection{General Ethical Principles}

In the United States, the \textit{Belmont Report} and the subsequent legislation in form of the \textit{Common Rule}, in particular the introduction of \textit{Institutional Review Boards} (IRBs), determined the ethics of government-funded research involving humans for more than 30 years~\cite{Vitak16}.
The Belmont Report contains three guiding ethical principles: respect for research participants, beneficence, and justice in participant selection~\cite{FederalRegister79}.
Respect for research participants means protecting their autonomy---they must enter the research ``voluntarily and with adequate information'' (informed consent).
For beneficence, two basic rules have been defined: (1) ``do not harm'' and (2) ``maximize possible benefits and minimize possible harms''.
Justice in participant selection implies a ``fairness in distribution'' of the burdens and benefits of research. 
Even if these principles are challenged by new developments like collecting and analyzing online data, they are still an important guideline for researchers~\cite{Vitak16, Shilton16}.

\subsection{The CASRO Code of Ethics }

To look out on how other communities handle ethical questions, we will now present an established code of standards and ethics, provided by the \emph{Council of American Survey Research Organizations} (CASRO)~\cite{Casro16}.
The CASRO code of ethics has a dedicated section about ``internet research''. 
One central statement in this section is that ``survey research organizations [must] not use unsolicited emails to recruit survey respondents or engage in surreptitious data collection methods''.
Further, researchers are required to ``verify that individuals contacted for research by email have a reasonable expectation that they will receive email contact for research''.
The CASRO code of ethics clearly defines when a researcher can assume that this is the case:
(i.) a substantive pre-existing relationship must exist with the contacted individual;
(ii.) the person receiving an email invitation has, based on the existing relationship, a reasonable expectation to be contacted for research purposes and he or she has not opted out for email communications;
(iii.) participants must not be recruited via unsolicited email invitations.

Moreover, for obtaining email addresses of potential participants, researchers must not collect ``email addresses from public domains'' and  use ``technologies or techniques to collect email addresses without individuals' awareness''.
In the following, we will now briefly look at the sampling strategies presented in Section~\ref{sec:strategies} and evaluate them according to the criteria from the CASRO code of ethics and the three ethical principles defined in the Belmont Report.

\subsection{Ethics of Sampling Strategies}
\label{sec:ethics-sampling}

When researchers use their personal network to recruit participants for their survey, in most cases it should be safe to assume that the CASRO criteria for ethical research are fulfilled.
The same applies when approaching companies using a gatekeeper.
However, when contacting companies to which no ``substantive pre-existing relationship'' exists, the first criterion is contravened.  
Using public media to reach developers would also be in line with the criteria, because participants are not directly contacted but read the call for participation on a news site and can then decide if they want to participate.
The sampling approach using data from GHTorrent would clearly contravene the criteria as well as the statement regarding the collection of email addresses from public domains.

In March 2016, users' email addresses were removed from the GHTorrent data dump after a heated debate on GHTorrent's issue tracker on GitHub about legal and privacy concerns raised by certain users~(see \cite{Ghtorrent16, Gousios16}).
This shows how sensitive the topic is and that a discussion in the research community is needed.
Beside the discussion on GitHub, there is also a discussion on StackExchange Academia about the ``Ethics of scraping 'public' data sources to obtain email addresses''~\cite{StackExchange16}, where the CASRO code of ethics is cited in the highest-ranked answer.
An important aspects of this discussion is what Brown et al. call a ``contextual concern''---researchers need to consider in what context users shared information online~\cite{Brown16}.
This also applies for GitHub, where users provide their email address, for instance, ``so people can contact [them] privately about problems in the community''~\cite{Ghtorrent16} and not to be contacted by researchers.

Approaching active GitHub developers using the email addresses they published on the platform also affects the three principles of the Belmont report: As more active users are likely to be contacted more often by researchers, justice in participant selection is not ensured. Further, beneficence is affected if developers change their behavior on GitHub in response to emails they perceive as spam (e.g. removing their email address from their profile page).
Also, the benefits of the research are reduced if the sampling strategy leads to biased samples. 
Lastly, depending on what other GitHub data researchers use, lack of informed consent may affect the principle of respect for participants.
El Emam points to the general problem that developers in open source projects probably never intended their work to be used for research projects~\cite{ElEmam01}.
Shilton and Sayles highlight the fact that much of the data available in social networks may have required informed consent for data collection in other settings~\cite{Shilton16}.
The same applies for data from GitHub and in particular the data available through GHTorrent.

Our intention is not to judge the research or the sampling practice that has been done in the past, but we want to start a discussion that may lead to new ethical guidelines for SE researchers.
We also do not want to promote the adoption of standards from other communities---in particular the CASRO code of ethics---but we think that they could inspire ethical guidelines for SE research.
In the discussion, we should also consider the 2012 report of the \textit{Association of Internet Research} (AoIR), which advocates flexible, process-oriented, and case-based guidelines instead of fixed code ethics~\cite{Markham12}.

Regarding the form of discussion, we can learn from the CSCW community, which organized several workshops and panels on ethical questions in the past~\cite{Fisher10, Fiesler15, Zytko15}, the computational social sciences community which published a dedicated book on ``Ethical Reasoning in Big Data''~\cite{Collmann16}, and the CHI community which discussed ethical issues at their 2016 conference~\cite{Brown16, Waern16}.
There exists some prior work in the SE community about research ethics~\cite{Wright06, Vinson08}, but since Vinson and Singer's statement in the year 2008 that ``the [empirical software engineering] community has yet to develop its own code of research ethics''~\cite{Vinson08}, not much has changed.

\begin{verticalline}
\textbf{Issue:} Compared to other communities, discussing ethical questions is not very common in the SE research community. Nevertheless, several sampling strategies employed by SE researchers in the past raise ethical concerns.\\
\textbf{Proposition:} We need a discussion at SE conferences and in program committees about ethical research practices. We can learn from other communities like CSCW and CHI and existing codes of ethics from other disciplines.
\end{verticalline}

\section{Conclusion}

In this paper, we reported on our experience with different approaches for sampling software developers.
The most efficient and effective strategies were to use public media and ``testimonials'' who advertise the survey.
We also highlighted the importance of gatekeepers who provide access to companies or communities.

Samples of software developers are often drawn in a non-random manner.
To be able the assess the external validity of studies involving such samples, researchers need a collection of typical demographics about software developers, which currently does not exist.
Using a systematic literature review, one could collect published demographics about developers.
Further, authors of studies with software developers could be contacted and asked to provide basic demographic information about their participants, if available.
This information, together with the data from the Stack Overflow Developer Surveys, would be a solid basis to assess the external validity of future studies.
Conferences and journals may recommend authors to describe certain key demographics for published studies and reviewers could motivate authors to explicitly address their sampling approach and effects on the generalizability of their results.

We also pointed at ethical issues with some of the sampling techniques researchers currently employ.
As mentioned above, we do not want to judge existing research and sampling techniques, but we want to start a discussion that may lead to new ethical guidelines for software engineering researchers.

\section{Future Work}

We plan to do a systematic literature review to collect published demographics about developers, employed sampling strategies, and reported sample sizes.
Caine conducted a similar literature review for samples of user studies in the CHI community~\cite{Caine16}.
The data from such a review could be the starting point for a database which other researchers can use to compare their sample to, as described above.
One challenge for this data set could be the different scales and categories for demographics such as work experience or software development roles.
However, in most cases it should be possible to adjust and compare the corresponding scales.
Further, an analysis of the publication culture regarding what demographics are reported and how sampling and potential biases are described may lead to recommendations for researchers.
Beside this literature review, we want to do a survey with software engineering researchers to answer the following research questions, adapted from Shilton and Sayles~\cite{Shilton16} and Vitak et al.~\cite{Vitak16}: RQ1: What do software engineering researchers believe constitutes ethical research?; RQ2: What are research ethics practices of software engineering researchers working with online datasets?; RQ3: What resources, i.e., codes of conduct, ethical guidelines, IRBs, do software engineering researchers consult?



%
\bibliographystyle{abbrv}
\bibliography{literature}  
%
%

\end{document}